\documentclass[12pt]{article}
\usepackage{graphicx}
\begin{document}

\title{ THE TRUMP PHENOMENON \\AN EXPLANATION FROM SOCIOPHYSICS}

\author{SERGE GALAM\thanks{serge.galam@sciencespo.fr} \\ CEVIPOF - Centre for Political Research, \\ Sciences Po and CNRS,\\
 98 rue de l'Universit\'e Paris, 75007, France}
\date{August 22, 2016}
\maketitle

\begin{abstract}

The Trump phenomenon is argued to depart from current populist rise in Europe. According to a model of opinion dynamics from sociophysics the machinery of Trump's amazing success obeys well-defined counter-intuitive rules. Therefore, his success was in principle predictable from the start. The model uses local majority rule arguments and obeys a threshold dynamics. The associated tipping points are found to depend on the leading collective beliefs, cognitive biases and prejudices of the social group which undertakes the public debate. And here comes the sesame of the Trump campaign, which develops along two successive steps.  During a first moment, Trump's  statement produces a majority of voters against him. But at the same time, according to the model the shocking character of the statement modifies the prejudice balance. In case the prejudice is present even being frozen among voters, the tipping point is lowered at Trump's benefit. Nevertheless, although the tipping point has been lowered by the activation of  frozen prejudices it is instrumental to preserve enough support from openly prejudiced people to be above the threshold. Then, as infuriated voters launch intense debate, occurrence of ties will drive progressively hostile people to shift their voting intention without needing to endorse the statement which has infuriated them. The on going debate does drive towards a majority for Trump. The possible Trump victory at November Presidential election is discussed. In particular, the model shows that  to eventually win the Presidential election, Trump must not modify his past shocking attitude but to appeal to a different spectrum of frozen prejudices, which are common to both Democrats and Republicans.

\end{abstract}

Keywords: Donald Trump; US presidential election; sociophysics.

\section{Introduction}

The current American campaign for both Democratic and Republican presidential nominations has gone through a series of successive unexpected outcomes. Not many  experts and analysts had imagined that Bernie Sanders would do so well against Hillary Clinton. Even less would have bet on Donald Trump nomination till the last moment before the Republican National Convention. 

When the Republican Party primaries for the 2016 United states presidential election started, most analysts were explaining with very sound arguments that Trump candidacy was a volatile phenomenon, which will very soon collapse. 

Indeed,  Trump nomination by the Republican party on July 22, 2016 has come as a total disclaimer to all analysts, who had kept predicting the eventual collapse of  Trump, election after election during the primaries. It was supposed to be a temporal phenomenon and accordingly should have faded away at some point during the campaign. View as a political bubble, before almost each primary or caucus election, commentators were prompt to announce its coming burst. Instead, Trump kept on winning delegates till reaching the official nomination, turning previous predictions as pure wishful thinking. 

After having gained the Republican nomination, Trump could well dismiss again most predictions by defeating Clinton on November election. Yet, his first success still needs a rational explanation besides denoting it as either an improbable accident or the expression of a rejection of the elite agenda. Many questions need to be answered like the following ones:

(i) How come Trump has succeeded in gaining so many states in contradiction with solid and sound analyses? 

(ii) How come Trump has defied pundits and usurped the Grand Old Party (GOP)  establishment?

(iii) How come Trump has succeeded to get 16 starting competitors to drop out the campaign?

(iiii) How come Trump made his most serious rival Ted Cruz to give up the race for nomination?

All these questions are linked to the main puzzle of Trump campaign: How come Trump kept infuriating and outraging a large spectrum of Republican voters by iterating a series of  ÒunacceptableÓ shocking statements and at the same time he got a majority of these people to eventually vote for him? It is this paradoxical singularity which has to be elucidated to understand the rationale behind Trump phenomenon.

To solve this enigma, I appeal to a model of opinion dynamics I have been developing for more than thirty years within the frame of so-called sociophysics \cite{book,review}. Denoted the Local Majority Rule (LMR) or the Majority Rule (MJ) or the Sequential Majority Rule (SMR) or the Galam model \cite{voting,chopa,mino,refe,hetero}. It articulates around three main points, which will be enumerated in next Section.

The LMR model is found to obey a threshold dynamics. For two competing opinions, the simplest case of homogeneous agents exhibits two attractors separated by a tipping point. When all discussing groups are odd, i.e., a local majority always exists and the tipping point is perfectly democratic, located at fifty percent. However in case of even size groups, breaking the tie at the benefit of one opinion suppresses the democratic balance at the advantage of that opinion. The associated  tipping points can go down as low as 15\% to 25\% for the opinion in tune with the leading belief while jumping at 85\% or 75\% for the competing other opinion.  

These threshold values are found to depend on the current distribution of collective beliefs and cognitive biases of the social group which undertakes the public debate. The question of which are the leading beliefs and prejudices activated in case of a local doubt is instrumental to understand the machinery of Trump amazing dynamics. There lies the key power of Trump campaign with his capacity to modify the current balance of active prejudices at his benefit.

Applying the LMR model abruptly to the American campaign for the Republican nomination would have yielded a priori the conclusion that Donald Trump candidacy will be defeated by more traditional Republican candidates like for instance Cruz, in accordance to most analyses. However, by making his repeated shocking statements Trump succeeded to accomplish two contradictory outcomes. The first one is visible and concerns the truly massive outrage produced among a good part of Republican voters. This effect results in producing a majority against Trump preserving only a small minority of support by people who do share the statement content. The second outcome concerns the leading prejudice among a majority of Republican voters. 

By outraging people, Trump awakes at the same time frozen prejudices which are connected to the content of indignation. On this ground, infuriated people who condemn Trump statement want to argue with others to affirm and legitimize their stand. However, they engage in a public debate being unaware of the activation of a frozen prejudice they indeed truly reject but only conscientiously standing against the associated statement. This is an instrumental hypothesis of our analysis. On this ground, once the debate is launched, people start to shift opinion with an eventual reversing of majority in favor of Trump provided his rescaled initial support value is above the threshold value. This initial support is made of the small minority of openly prejudiced people.

The LMR model provides a clear ground to explain such a counter-intuitive mechanism. Therefore, Trump's success was in principle predictable from the start. Accordingly, Trump phenomenon departs from current populist rise in Europe like with Marine Le Pen of the National Front in France and Nigel Farage of the UK Independence party. It is a singular phenomenon which relies on the existence of frozen prejudices which are well locked by the people who share them against their conscious will.

The methodological approach used in this article subscribes to the novel sociophysics trend, which is developing among physicists at the level of international research \cite{santo,bikas,bolek,cheon,weron-t,celia,koree,gerard,nuno,bagnoli,carbone,weron,zanette,iglesias,mobilla,fasano,andre,marcel,marco}. Recently mathematicians are also getting involved in the mathematical modeling of social phenomena \cite{bello,lanchier}. Sociophysics has been also applied to model terrorism \cite{terro1,terro2,terro3,podo} and war \cite{sen1,sen2}. While sociophysics is today a flourishing field, it was strongly criticized among physicists at its earlier steps \cite{testi}.

It is worth to stress that while sociophysics is still at its earlier stage of construction, some real world events have been successfully predicted. Among others, voting outcomes like the 2005 French No victory for the European constitution referendum \cite{refe} the 2002 victory of JM LePen at the first run of French presidential election \cite{libe} as well as the repetition of fifty-fifty votes in democratic societies following the 2000 Bush-Gore election in the US \cite{fifty}.

More recently the Brexit vote which came as a total blow to all analysts, was envisioned in 2004 using the LMR model \cite{refe} with the following concluding statement:
\begin{quote}
{\it ``Applying our results to the European Union leads to the conclusion that it would be rather misleading to initiate large public debates in most of the involved countries. Indeed, even starting from a huge initial majority of people in favor of the European Union, an open and free debate would lead to the creation of huge majority hostile to the European Union. This provides a strong ground to legitimize the on-going reluctance of most European governments to hold referendum on associated issues."}
\end{quote}

The rest of the paper is organized as follows. Next Section presents the Local Majority Rule (LMR) model of opinion dynamics.Tipping points are discussed in Section 3 while Section 4 presents a general illustration of the dynamics which exhibits counter intuitive results. The Trump paradoxical sesame is given an instrumental explanation in Section 5. The question of whether or not Trump will be the next US president at the November, 2016 election, is given some answer in Section 6. In particular, the model shows that  to eventually win the Presidential election Trump does not need to modify his past shocking attitude but to appeal to a different spectrum of frozen prejudices, which are common to both Democrats and Republicans.The conclusion contains several clarifying statements about the implications of this work and the meaning of making an either successful or wrong prediction.

\section{The Local Majority Rule Model}

The LMR model articulates around three successive steps, which are iterated several times in a row.

\begin{enumerate}
\item First step: distribute the agents randomly in different groups of small sizes from 1 up to an upper limit usually equal to 5 or 6.
\item  Second step: apply a local majority rule within each single group to update all agent choices along the choice which got the majority of votes. In case of an even size group at a tie update along one choice with probability $k$ and the other with probability $(1-k)$. The current value of $k$ is  a function of the actual leading prejudice and belief within the associated social group.
\item Third step: reshuffle all the agents and restart from step 1.
\end{enumerate}

For situations where two choices A and B are competing with respective initial supports $p_0$ and  $(1-p_0)$, one series of steps 1-2-3 shifts $p_0$ to a new value $p_1$. After $n$ iterations, we get a proportion $p_n$. In order to evaluate quantitatively the successive changes of respective supports for A and B we calculate $p_1$ as
\begin{equation}
p_1\equiv P_T(p_0)= \sum_{r=1}^{L} a_r P_r(p_0) ,
\label{pT} 
\end{equation}
where $a_r$ is the proportion of groups of size $r$ with the constraint $\sum_{r=1}^{L} a_r=1$ and $P_r$ is the probability to have a group of $r$ agents (size $r$) being updated along choice A. L is the largest size of local groups, usually around 5 or 6. Larger groups tend to fragment spontaneously into smaller subgroups. More precisely, for any size $r$, both odd and even, the voting function $P_r$ writes
\begin{eqnarray}
P_r(p_0)&=& \sum_{m=N[ \frac{r}{2} ]}^{r}  {r \choose m} p_0^m  (1-p_0)^{r-m}
\nonumber\\
& +& k \delta \Bigg [ \frac{r}{2}-N\left [\frac{r}{2}\right ] \Bigg ] {r \choose \frac{r}{2}} p_0^\frac{r}{2}  (1-p_0)^\frac{r}{2},
\label{pr}
\end{eqnarray}
where ${r \choose m}\equiv \frac{r!}{m!(r-m)!}$, $k$ is an integer with $0\leq k \leq 1$, $N[x]\equiv Integer\  part$ of  $x$ and $\delta [x]$ is the Kronecker function, i.e., $\delta [x]=1$ if $x=0$ and $\delta [x]=0$ if $x\neq 0$.

First term in Eq. (\ref{pr}) accounts for local majorities with  configurations $\{r A, 0 B\}$, $\{(r-1) A, 1 B\}$, $\{(r-2) A, 2 B\}$... , $\{(N\left [\frac{r}{2}\right ] +1) A, (N\left [\frac{r}{2}\right ] -1) B\}$ while second term contribution is produced from a tie at an even size group with $\{\frac{r}{2} A, \frac{r}{2} B\}$. 

There, the factor $k$ breaks the symmetry between both choices. At a tie, with as many arguments for one choice as for the other, the group get trapped into a collective doubt. In this situation all agents chose A with probability $k$  and B with probability $(1-k)$.  Only $k=\frac{1}{2}$ restores the symmetry between A and B.

It is worth to stress that any odd size $r$ yields $\{\frac{r}{2}-N[\frac{r}{2}]\}=\frac{1}{2}$ making the Kronecker function always equals to zero in Eq. (\ref{pr}) thus canceling the second  term.  For even $r$ size $\{\frac{r}{2}-N[\frac{r}{2}]\}=0$ with last term equal to $k {r \choose \frac{r}{2}} p_0^\frac{r}{2}  (1-p_0)^\frac{r}{2}$. 

Eq. (\ref{pr}) can thus simply be written in two different expressions
\begin{equation}
 P_r(p_0) \equiv  \sum_{m= \frac{r+1}{2} }^{r}  {r \choose m} p_0^m  (1-p_0)^{r-m} ,
\label{pr-odd} 
\end{equation}
for odd sizes $r$, and
\begin{equation}
 P_r(p_0)\equiv \sum_{m= \frac{r}{2} +1}^{r}  {r \choose m} p_0^m  (1-p_0)^{r-m}
+k {r \choose \frac{r}{2}} p_0^\frac{r}{2}  (1-p_0)^\frac{r}{2},
\label{pr-even} 
\end{equation}
when $r$ is even. Note that the summation term is identical to Eq. (\ref{pr-odd}) besides for the summation first term which is $ \frac{r}{2} +1$ instead of $\frac{r+1}{2}$ for the odd case.

In the simplest case $r=3$, Eq. (\ref{pr-odd}) writes
\begin{equation}
P_3(p_0)=p_0^3+3 p_0^2 (1-p_0) ,
\label{p3} 
\end{equation}
which accounts for all local $A$ majority within a group of 3 agents $\{3 A , 0 B\}$ and $\{2 A, 1 B\}$. The multiplicative factor 3 in last term accounts for the three possibilities to place  one B among two A. 

In contrast, $r=4$ yields from Eq. (\ref{pr-even})
\begin{equation}
P_4(p_0)=p_0^4+4 p_0^3 (1-p_0) +6 k p_0^2 (1-p_0)^2 ,
\label{p4} 
\end{equation}
which includes both local majorities with $\{4 A , 0 B\}$ and $\{3 A, 1 B\}$ and the tie $\{2 A , 2 B\}$.

\begin{figure}
\centering
\includegraphics[width=.80\textwidth]{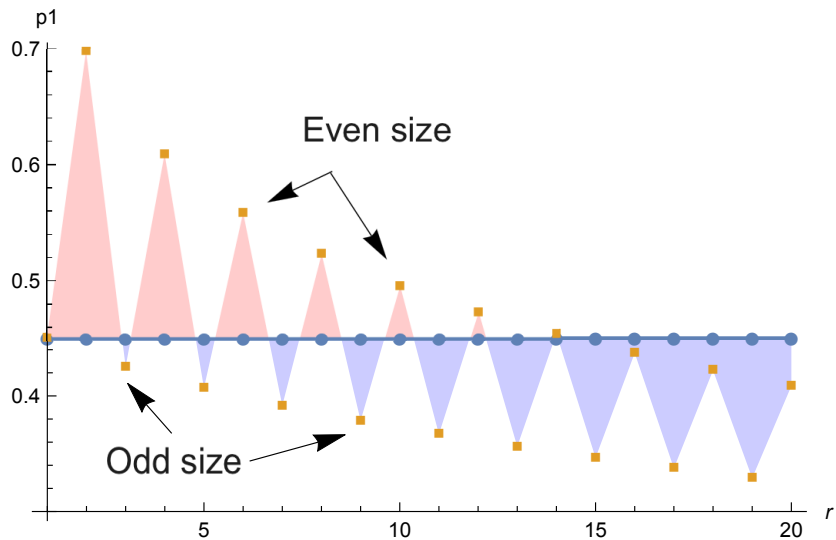}
\caption{Given $p_0=0.45$ respective values of $p_1$ are shown for $r=1,2,3, ...,20$ in case $k=1$.
Odd sizes yield always values lower than $0.45$. The larger the size, the larger the amplitude from $0.45$. In contrast, even sizes up to $r=14$ yield values higher then $0.45$. Contrary to odd sizes, the larger the size, the smaller the amplitude from $0.45$. From $r=16$ and up the behavior is similar to odd sizes.}
\label{p1-r}
\end{figure}

For instance, for $p_0=0.450>0.500$ Eqs. (\ref{pr-odd})  and (\ref{pr-even}) yield  $p_1=0.203 + 0.495 k, 0.425, 0.241 + 0.368 k$, $0.407, 0.255 + 0.303 k$ for respectively $r=2,3,4,5,6$. The series turns to $p_1=0.698, 0.425, 0.609$,  $0.407, 0.558$ for $k=1$. Associated values are exhibited in Figure (\ref{p1-r}). 

\begin{figure}
\centering
\includegraphics[width=.80\textwidth]{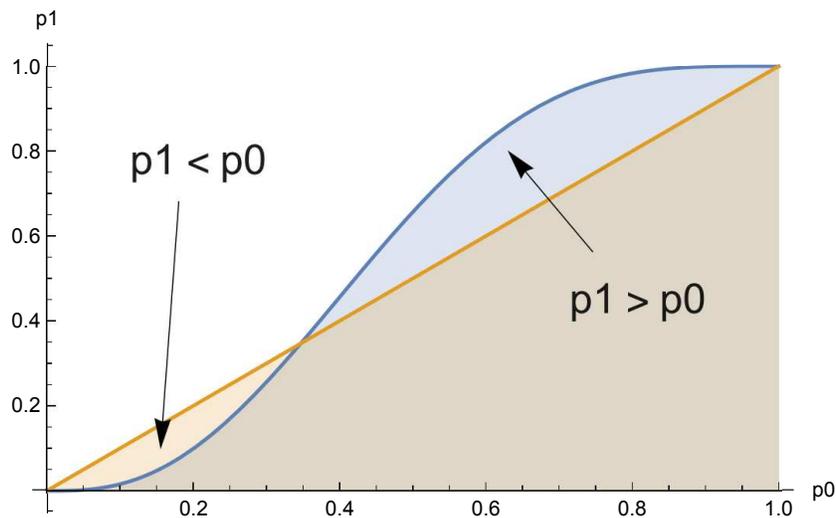}
\caption{The vlaue $p_1$ is shown as a function of $p_0$ in the full range $0\leq p_0 \leq 1$ for $k=1$ and $r=4$.}
\label{p1-p-r=4}
\end{figure}

It is seen that odd sizes yield always values lower than $0.45$. The larger the size, the larger the amplitude from $0.45$. In contrast, up to $r=14$ even sizes yield values higher then $0.45$. Contrary to odd sizes, the larger the size, the smaller the amplitude from $0.45$. From $r=16$ and up the behavior is similar to odd sizes. By reversed symmetry $k=0$ gives $p_1=0.203, 0.425$, $0.241, 0.407, 0.255$. Figure (\ref{p1-p-r=4}) shows  the behavior of $p_1$ at $k=1$ and $r=4$ as a function of $p_0$ in the full range $0\leq p_0 \leq 1$.

On the other side of the democratic threshold, $p_0=0.550 > 0.500$ gives $p_1=0.303 + 0.495 k, 0.575$, $0.391 + 0.368 k, 0.593$, $0.442 + 0.303 k$ for $r=2,3,4,5,6$. It turns to $p_1=0.798, 0.575, 0.759, 0.593, 0.745$ for $k=1$ against $p_1=0.303, 0.575, 0.391$, $0.593, 0.442$ when $k=0$.

Table (\ref{pr-1}) shows the various values of $p_1$ obtained for respectively $p_0=0.100, 0.200$, $0.300, 0.400, 0.450, 0.480$, $0.500, 0.520, 0.550$, $0.600, 0.700, 0.800, 0.900$ using for each value, $r=2,3,4,5,6$. Results are rounded at three digits. All these illustrations show clearly that odd sizes respect a democratic balance with always $p_1>p_0$ when $p_0>0.500$ and  $p_1<p_0$ when $p_0<0.500$. In contrast, even size can increase a minority value with $p_1>p_0$ for $p_0<0.500$ and  $p_1<p_0$ when $p_0>0.500$ depending on $k$ and $p_0$.

\begin{table}
\caption{The single group voting outcome $p_1$ is displayed for respectively $p_0=0.100, 0.200, 0.300, 0.400$, $0.450, 0.480$, $0.500, 0.520$, $0.550, 0.600, 0.700, 0.800, 0.900$ using $r=2,3,4,5,6$. Data are rounded at three digits.}
{\begin{tabular}{| c c c c c c | }
\hline\noalign{\smallskip}
$p_0 $  & $p_1,r=2$ & $p_1,r=3$ &  $p_1,r=4$ &  $p_1,r=5$  &  $p_1,r=6$ \\ \noalign{\smallskip}\hline\noalign{\smallskip}
0.100& 0.010 + 0.180 k& 0.028& 0.004 + 0.049 k& 0.009& 0.001 + 0.015 k\\ \noalign{\smallskip}\hline\noalign{\smallskip} 
0.200& 0.040 + 0.320 k& 0.104& 0.027 + 0.154 k& 0.058& 0.017 + 0.082 k\\ \noalign{\smallskip}\hline\noalign{\smallskip} 
0.300& 0.090 + 0.420 k& 0.216& 0.084 + 0.265 k& 0.163& 0.070 + 0.185 k\\ \noalign{\smallskip}\hline\noalign{\smallskip} 
0.400& 0.160 + 0.480 k& 0.352& 0.179 + 0.346 k& 0.317& 0.179 +0.276 k\\ \noalign{\smallskip}\hline\hline\noalign{\smallskip} 
0.450& 0.203 + 0.495 k& 0.425& 0.241 + 0.368 k& 0.407& 0.255 + 0.303 k\\ \noalign{\smallskip}\hline\noalign{\smallskip} 
0.480& 0.230 + 0.499 k& 0.470& 0.283 + 0.374 k& 0.463& 0.307 + 0.311 k\\ \noalign{\smallskip}\hline\noalign{\smallskip} 
0.500& 0.250 + 0.500 k& 0.500& 0.313 + 0.375 k& 0.500& 0.344 + 0.313 k\\ \noalign{\smallskip}\hline\noalign{\smallskip} 
0.520& 0.270 + 0.499 k& 0.530& 0.343 + 0.374 k& 0.537& 0.382 + 0.311 k\\ \noalign{\smallskip}\hline\noalign{\smallskip} 
0.550& 0.303 + 0.495 k& 0.575& 0.391 + 0.368 k& 0.593& 0.442 + 0.303 k\\ \noalign{\smallskip}\hline\hline \noalign{\smallskip} 
0.600& 0.360 + 0.480 k& 0.648& 0.475 + 0.346 k& 0.683& 0.544 + 0.276 k\\ \noalign{\smallskip}\hline\noalign{\smallskip} 
0.700& 0.490 + 0.420 k& 0.784& 0.652 + 0.265 k& 0.837& 0.744 + 0.185 k\\ \noalign{\smallskip}\hline\noalign{\smallskip} 
0.800& 0.640 + 0.320 k& 0.896& 0.819 + 0.154 k& 0.942& 0.901 + 0.082 k\\ \noalign{\smallskip}\hline\noalign{\smallskip} 
0.900& 0.810 + 0.180 k& 0.972& 0.948 + 0.049 k& 0.991& 0.984 + 0.015 k \\ \noalign{\smallskip}\hline\noalign{\smallskip} 
\end{tabular}}
\label{pr-1}  
\end{table}

\section{Tipping points dynamics}

Having calculated $p_1$ from $p_0$ using Eq. (\ref{pT}) we need to iterate the update to account for an on going debate about a given issue. It means to calculate successively $p_2=P_T(p_1)$, $p_3=P_T(p_2)$, ..., $p_{n}=P_T(p_{n-1})$ for $n$ consecutive updates of opinions. Each update takes some amount of real time, which depends on the intensity of the ongoing debate. The value of $n$ is scaled with the overall debate duration.  In case of a vote the value of $n$ can be determined using a series of polls and the campaign duration from its launch till the actual voting. At the voting what counts for A is to have $p_n>\frac{1}{2}$ to win the election and $p_n<\frac{1}{2}$ for B to win.

To apprehend the associated dynamics and be able to make a prediction ahead of the actual voting we need to study the attractor landscape which underlines the dynamics of $P_T(p_0)$ given by Eq. (\ref{pT}). Accordingly we solve the so-called fixed-points equation
\begin{equation}
P_T(p)=p ,
\label{pc} 
\end{equation}
which always satisfies $P_T(0)=0$ and $P_T(1)=1$ as could be expected. No spontaneous formation of choices is included in the model. To determine the respective stabilities of these two fixed points $p_B=0$ and  $p_A=1$ we need to calculate 
\begin{equation}
\lambda_T (p) \equiv \frac{dP_T(p)}{dp}|_{p} .
\label{dp} 
\end{equation}
The fixed point is stable when $\lambda_T (p) <1$ and unstable for $\lambda_T (p) >1$. Eq. (\ref{dp}) gives
\begin{equation}
\lambda_r(p)= \sum_{m=\frac{r+1}{2}}^{r} (m -r p) {r \choose m}  p^{m-1} (1-p)^{r-m-1} \ ,
\end{equation}
for odd sizes (Eq. (\ref{pr-odd})) and
\begin{eqnarray}
\lambda_r(p)&=&\sum_{m=\frac{r}{2}+1}^{r} (m -r p) {r \choose m}  p^{m-1} (1-p)^{r-m-1} \nonumber\\
& +& k (1-2p) \frac{r}{2} {r \choose \frac{r}{2}} p^{\frac{r}{2}-1} (1-p)^{\frac{r}{2}-1}  \ ,
\label{}
\end{eqnarray}
for even sizes (Eq. (\ref{pr-even})).

For both odd and even sizes we always have $\lambda_T (0) =\lambda_T (1)=0$, which shows that both fixed points are stable. They are the attractors of the opinion dynamics, which thus implies the existence of a tipping point, i.e., an unstable fixed point $p_c$ in between them.

Eq. (\ref{pc}) is a polynomial equation of degree $L$ as seen from Eq. (\ref{pT}). It thus have $L$ roots of which two have been identified as $p_B=0$ and  $p_A=1$. It is worth to stress that not every root is relevant to the actual opinion dynamics. Indeed, two constraints must be satisfied with roots being real and located between 0 and 1. 

While a general analytical solving seems to be out of reach, at least for me, restraining to odd sizes with Eq. (\ref{pr-odd}) shows that $p_c=\frac{1}{2}$ is always a fixed point. The associated stability is given by
\begin{equation}
\lambda_r(p_c)=\frac{1}{2^{r-1}}
\sum_{m=\frac{r+1}{2}}^{r} (2 m -r) {r \choose m}  \ ,
\end{equation}
which can be reduced to
\begin{equation}
\lambda_r(p_c) =\frac{r}{2^{r-1}}  {r-1 \choose \frac{r-1}{2}} \ .
\label{lambda-1/2}
\end{equation}
Eq. (\ref{lambda-1/2}) yields $\lambda_r(p_c)>1$ for any size with $\lambda_3(p_c)=\frac{3}{2}$ at $r=3$ to $\lambda_r(p_c) \rightarrow \sqrt \frac{2}{\pi} \frac{r}{\sqrt {r-\frac{2}{3}}}   \approx  \sqrt \frac{2r}{\pi} $ for 
$r \gg 1$ as seen in Table (\ref{lambda}). 

\begin{table}
\caption{ Variation of $\lambda_r(p_c)$ as function the local group  size with  $r=3, 5, 7, 9,15, 101, 1001$. It is always larger than 1 making the fixed point $p_c=\frac{1}{2}$ unstable.}
{\begin{tabular}{| c c c c c c c c | }
\hline\noalign{\smallskip}
 r & $ 3$ & $ 5$ &  $ 7$ &  $ 9$ &  $ 15$ & $ 101$ &  $ 1001$ \\
\noalign{\smallskip}\hline\noalign{\smallskip}
$\lambda_r(p_c)$ & 1.5 & 1.87 & 2.19 & 2.46 & 3.14 & 8.04 & 25.25 \\ \hline
\end{tabular}}
\label{lambda}   
\end{table}

Therefore, $p_c=\frac{1}{2}$ is the tipping point of an opinion dynamics restricted to discussing groups of odd size. Indeed, the result holds true for even sizes provided $k=\frac{1}{2}$, which is a special case. Otherwise, when $k\neq \frac{1}{2}$, $p_c=\frac{1}{2}$ is not a fixed point for even sizes and Eq. (\ref{pc}) has to be solved numerically. The case $r=4$ can be solved analytically with
\begin{equation}
p_{c,4,k}=\frac{1 - 6 k + \sqrt{13 - 36 k + 36 k^2}}{6 (1 - 2 k)}\ ,
\label{pc4}
\end{equation}
which yields $p_{c,4,1}\approx 0.232$ and $p_{c,4,0}\approx 0.768$.

Accordingly, when $k=1$ opinion A needs to start with $p_0>0.232$ to end up being majority after a few updates as seen in  Figure (\ref{p1-p-r=4}). In contrast, for $k=0$ opinion A must have $p_0>0.768$ to preserve its majority status once a debate is launched. 

\begin{figure}
\centering
\includegraphics[width=.80\textwidth]{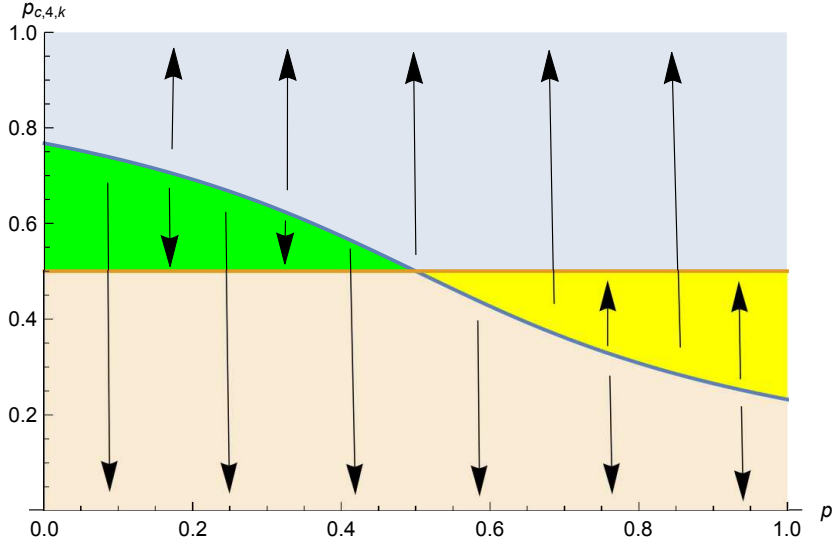}
\caption{The tipping point $p_{c,4,k}$ from Eq. (\ref{pc4}) as a function of $k$ in the full range $0\leq k \leq 1$. Arrows indicate the direction of the opinion drift as a result of repeated local discussions. For $\frac{1}{2}< k\leq1$ a full region exists where an initial A minority is turned majority thanks to the bias driven by the leading prejudices of the social group. By reversed symmetry when $0\leq k<\frac{1}{2}$ the leading prejudices are driven the opinion flow at the expense of A as seen from the arrows.}
\label{pc-4-k}
\end{figure}

Figure (\ref{pc-4-k}) shows the variation of  the tipping point $p_{c,4,k}$ from Eq. (\ref{pc4}) as a function the bias breaking $k$. For $\frac{1}{2}< k\leq1$ a full region exists where an initial A minority is turned majority thanks to the bias driven by the leading prejudices of the social group. The 
arrows indicate the direction of the opinion drift as a result of repeated local discussions. By reversed symmetry when $0\leq k<\frac{1}{2}$ the leading prejudices are driven the opinion flow at the expense of A  and the benefit of B.

\begin{figure}
\centering
\includegraphics[width=.80\textwidth]{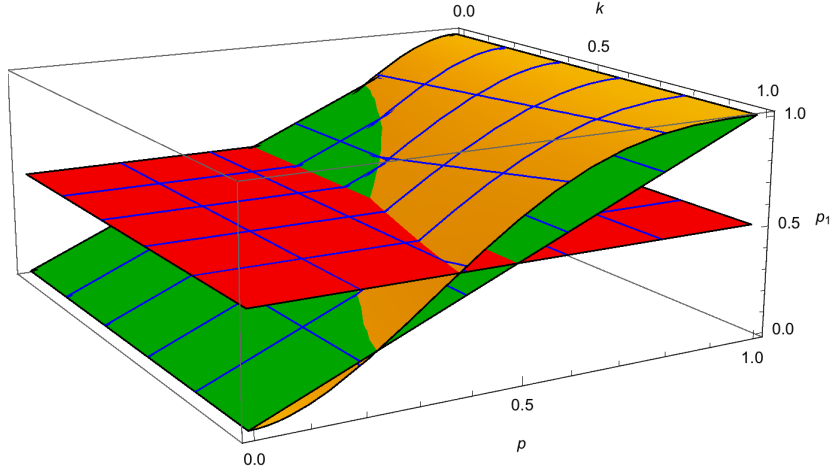}
\caption{The function $p_1=P_{4}$ from Eq. (\ref{p4}) as a function of $k$ and $p$ is exhibited with both planes $p_1=\frac{1}{2}$ and $p_1=p$. The intersection of last plane and $p_1=P_{4}$  determines the line of tipping points $p_{c,4,k}$. It start at $0.232$ at $k=1$ to end up at $0.768$ for $k=0$.}
\label{p1-4-k}
\end{figure}

\section{Counter intuitive dynamics: a general illustration}

To grasp fully the counter intuitive dynamics driven by democratic informal small groups debates we now illustrate the model through one general illustration with two cases. In the first case people are distributed within groups of size $r=1,2,3,4,5$ and only  $r=1,2,3,4$ in the second case. Proportions  are equal with $a_1=a_2=\dots = a_L=\frac{1}{L}$, i.e. $\frac{1}{5}=0.200$ and $\frac{1}{4}=0.250$ respectively. 

When $k=1$ associated tipping points are found to be $p_{c,r_{1\dots 5},1}=0.278$ and  $p_{c,r_{1\dots 4},1}=0.153$. To win the public debate and cross fifty percent at the election, opinion A must start from a minimum support $p_0>0.278$ in the first case and only $p_0>0.153$ in the second case. It is the reverse situation for opinion B with initial supports larger than $0.722$ and $0.847$. The situation appears to be far more favorable for opinion A against B, which has almost no chance to win the election once a debate has been taken place.

\begin{figure}
\centering
\includegraphics[width=1\textwidth]{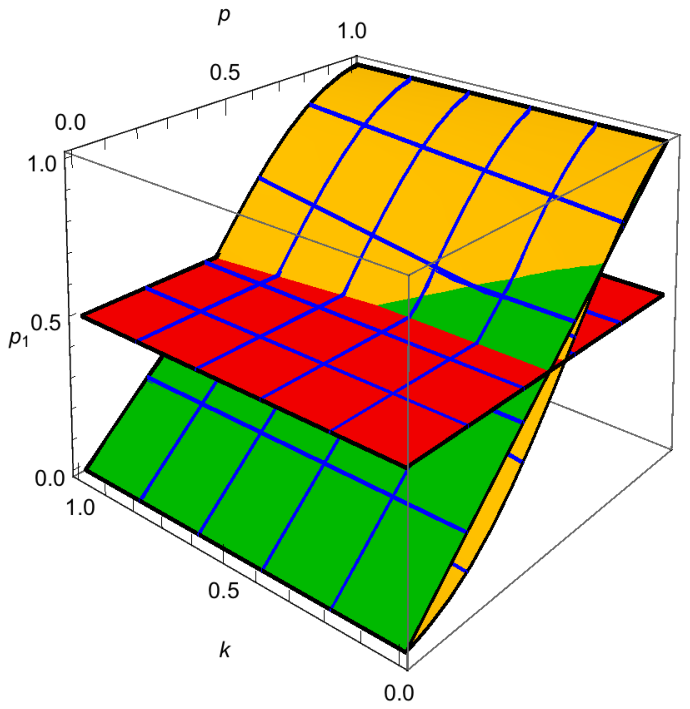}
\includegraphics[width=1\textwidth]{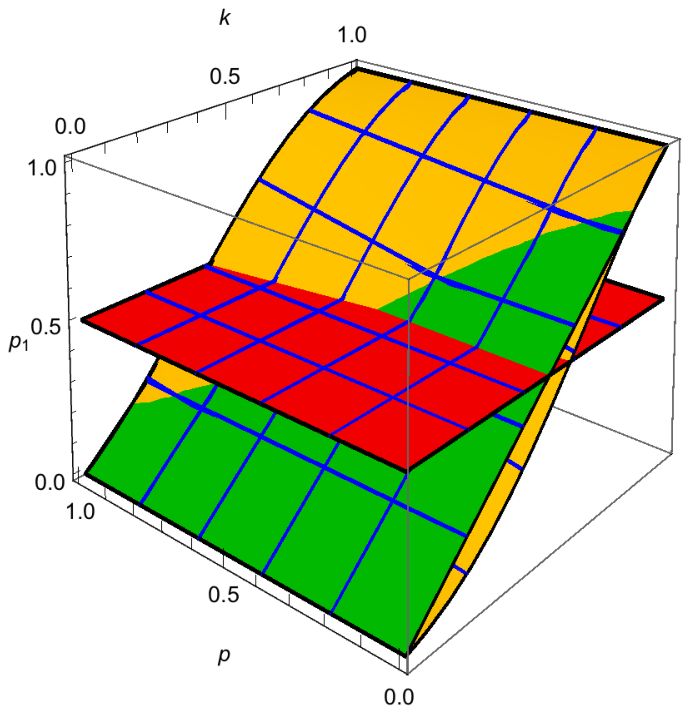}
\caption{Same as Figure  (\ref{p1-4-k}) for $p_1=P_{T}$ from Eq. (\ref{pT}) with $a_1=a_2=\dots = a_L=\frac{1}{L}$, i.e. $\frac{1}{5}=0.200$ (top) and $\frac{1}{4}=0.250$ (bottom). The lines of tipping points start respectively at $0.278$ and $0.153$ at $k=1$ to end up at $0.722$ and $0.847$ for $k=0$.}
\label{g1-2}
\end{figure}

Accordingly an opinion which starts with a support below its tipping point will eventually lose support along the on going informal discussions among people.  The instrumental question being then to determine if this opinion is in tune or in contradiction with the current leading prejudice and belief shared by the associated social group in order to identify which of the two opposite  tipping points is concerned. 

At this stage it is importance to underline that the building of prejudice and belief among a social group obeys a rather slow dynamics making them fixed and constant during a public debate. The time rate of opinion dynamics is indeed much more quicker of the order of days in contrast to prejudice and belief for which typical time rate is of the order of years if not decades.

An opinion starting with a vey high support of individual choices is thus scheduled to lose this support as soon as the debate is launched in case it is in dissonance with the leading prejudice and belief. For instance $p_0=0.80$ yields the series $0.789,0.774,0.756,0.731,0.699,0.655,0.595,0.514,0.408$ for $9$ successive updates using the distribution $a_1=a_2=a_3=a_4=\frac{1}{4}=0.250$ at $k=0$. A huge majority is turned minority due to the contradiction of the choice with the leading prejudice and belief. 

Within the same group setting an initial minority associated to $k=0$ fades away quite rapidly as seen with  $p_0=0.45$  and the series $0.330,0.200$, $0.093,0.032$ for $4$ updates.

\section{The Trump paradoxical sesame}

Trump candidacy is in contradiction with the leading prejudices and beliefs of Republicans and Democrats as it was evidenced from all earlier analyses about Trump candidacy. 

Besides polarizing so called anti-system feeling Trump profiles did not comply with the usual basic skills requested to become both the nominee of the Republican party and the president of the United State. Therefore, according to my model, even if he could crystalize some initial majority support, this support should have weakened progressively making Trump lose many of the series of intermediate elections to eventually loose the Republican nomination. 

Such a faith is in agreement with most analyses, which had  concluded Trump was vowed to lose the run for Republican nomination. He was portrayed as a political bubble aimed at a rather quick collapse.

It is of importance to underline that among his Republican supporters many were asking him to reverse such a provocative path, which should necessarily make him lose.  Analysts would suggest to build a more respectable and experienced image to fit the standards of a solid candidate to the presidential run. 

And here come the rationale of Trump apparent irrational attitude. At least according to my analysis. Instead of trying to win voter support before most local election he kept on going frontally against the voter values getting them infuriated. He thus lost many of his supporters. 

At this moment the move sounds totally counter-productive and even suicidal. And indeed, the first outcome of such a move is effectively to erode substantially Trump support. However at this very moment the sesame of Trump phenomenon appears in its powerful mechanics deploying itself in a threefold process.
\begin{description}
\item[(i)] The process which gets people infuriated activates at the same time the underlying associated prejudice, which were present but frozen. This activation upsets the current balance of shared prejudice and common belief of the social group at the benefit of the newly activated belief. Within our framework, it turns $k$ from zero to one.
\item[(ii)] Being infuriated by a shocking statement, people want to prove to themselves and to their peers that  they reject the statement and thus do engage into a fierce  debate. By so doing the opinion dynamics is launched driven by the new leading prejudice, which is precisely the one people are departing from on a rational basis. They will not vote for Trump.
\item[(iii)] However within even size groups ties occur making the corresponding group to doubt about which choice to select. A collective doubt is built from a rational confrontation of opposite arguments. At this stage the group eventually does chose Trump driven by the unconscious activated prejudice, without having to formally identify with the prejudice. 
\end{description}

\begin{figure}
\centering
\includegraphics[width=.80\textwidth]{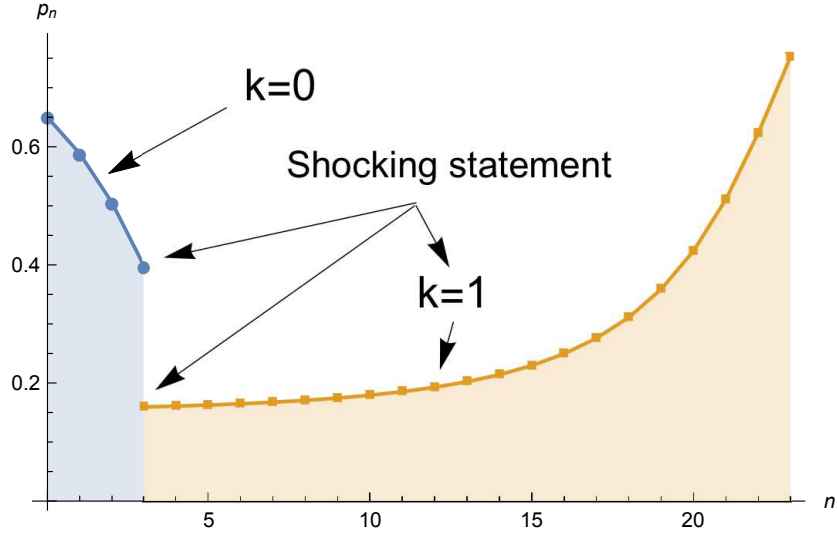}
\caption{Values $p_n=P_{T}(p_{n-1})$ using Eq. (\ref{pT}) for the case $L=4$, i.e., groups of size $r=1,2,3,4$ with equal proportions  $a_1=a_2=a_3= a_4=\frac{1}{4}=0.250$. Tipping points are $0.847, 0.153$ for respectively $k=0,1$. Here $p_0=0.65$. The first regime is with $k=0$. After 3 updates$p_3=0.397$. At this moment Trump makes a shocking statement which drives $p_3$ down to $0.160$. This shift is not produced by discussions but by individual changes of opinion. Simultaneously to the support decrease, the shocking statement  activates a prejudice associated to the statement nature and provokes a debate fueled by infuriated people. And here comes the "miracle" or the ''catastrophe" depending on people political side, the support starts to increases first slowly with $p_4=0.160$, $p_5=0.163$, $p_6=0.165, \dots$ and then accelerates with $p_{18}=0.311,\dots, p_{21}=0.511$, $p_{22}=0.623, p_{23}=0.752$.}
\label{t1}
\end{figure}

We illustrate this unanticipated spin-off of the shocking statement by considering an initial support $p_0=0.65$ as exhibited in Figure  (\ref{t1}). We assume without loss of generality the case of group distribution  $L=4$, i.e., groups of size $r=1,2,3,4$ with equal proportions  $a_1=a_2=a_3= a_4=\frac{1}{4}=0.250$. The dynamics is initiated with $k=0$ and leads to $p_1=0.588, p_2=0.505, p_3=0.397$ for 3 successive updates. The support weakens since the tipping point is located at $p_0<p_{c,r_{1\dots 4},0}=0.847$.

At this moment Trump makes a shocking statement and consequently lose a good part of his supporters now down let say to $p_3=0.160$. This rescaling of $p_3$ is driven by a series of individual reaction of supporters. It is not the result of discussions among agents. 

Once the support $p_3$ has been rescaled down, the balance of prejudice is also reset at Trump's benefit with $k=1$ instead of $k=0$. The debate resumes but since $p_3=0.160>p_{c,r_{1\dots 4},1}=0.153$, the support starts to grow. The increase is very slow during the first series of updates till some range where it starts jumping to cross the winning value of $50\%$ as seen from in Figure  (\ref{t1}).

Nevertheless, above reversing of opinion trend is not guaranteed for any configurations. Some cases can be sensitive to small support differences. For instance, consider the case $L=6$ with $a_1=0.10, a_2=0.30, a_3= 0.20, a_4=0.30,a_5=0.05,a_6=0.05$. It yields $p_{c,r_{1\dots 6},0}=0.837$ and $p_{c,r_{1\dots 6},1}=0.163$. Starting from the same initial support $p_0=0.65$ leads now to $p_1=0.575, p_2=0.465, p_3=0.315$ for 3 successive updates. Results are similar to above case.

\begin{figure}
\centering
\includegraphics[width=.50\textwidth]{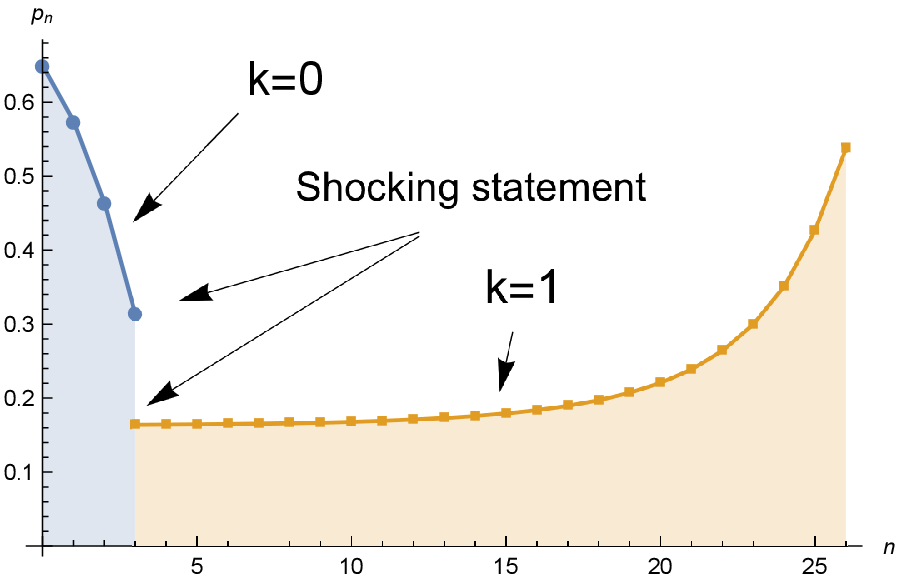}\includegraphics[width=.50\textwidth]{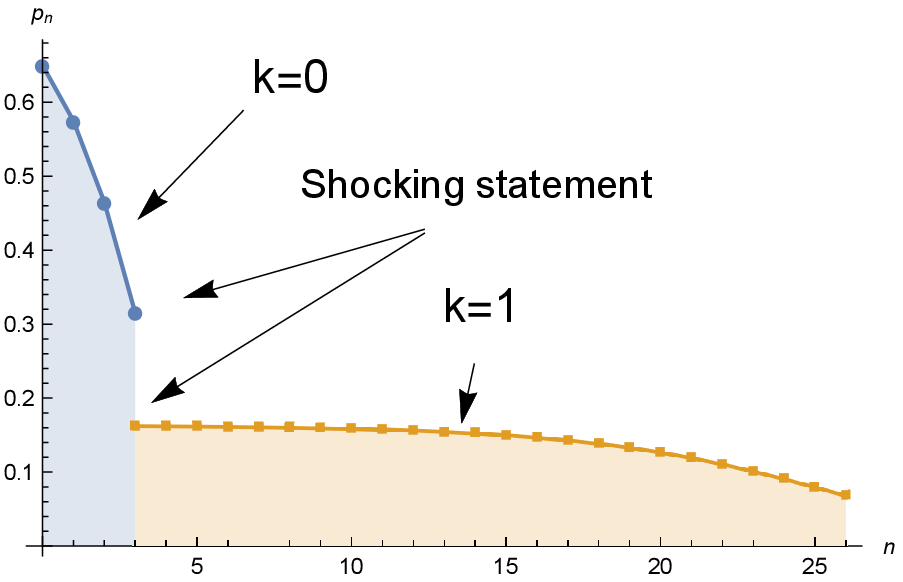}
\caption{Same as Figure (\ref{t1}) with small but critical changes. The discussion group distribution is given by $a_1=0.10, a_2=0.30, a_3= 0.20, a_4=0.30,a_5=0.05,a_6=0.05$ with $L=6$. Associated tipping points are $p_{c,r_{1\dots 6},0}=0.837$ and $p_{c,r_{1\dots 6},1}=0.163$. initial support is still $p_0=0.65$. Opinion dynamics is first taking place with $k=0$ leading after 3 updates to $p_1=0.575, p_2=0.465, p_3=0.315$. The shocking statement is made at the same moment as before but now rescaled down to $p_3=0.164$ in the left part and to $p_3=0.162$ in the right part. From there, opinion dynamics takes place with $k=1$ for both cases. While in the left part, the evolution is similar to that of Figure (\ref{t1}) since $p_3>p_{c,r_{1\dots 6},1}$ with Trump victory, it is opposite on the right part with $p_3<p_{c,r_{1\dots 6},1}$ and Trump defeat.}
\label{t2}
\end{figure}

Rescaling $p_3=0.315$ down to $p_3=0.164$ produces also a similar behavior as above. Results are shown in left part of Figure (\ref{t2}). However, if the falling down of support driven by the shocking statement is a little bit bigger down to $p_3=0.162$ the overall dynamics is put up side down as illustrated in the right part of Figure  (\ref{t2}). There, Trump does lose since his after statement support went below the tipping point. While the reversal process is overall quite robust, it could turn fragile in the vicinity of the tipping point. 

\section{Will Trump be the next US president?}

Not too many had foreseen Donald Trump gaining the Republican party nomination and the same  vision is prevailing for the November, 2016 US Presidential election. 
Strengthened by polls giving a substantial gap in favor to Hillary Clinton, most analysts reiterate their previous wrong forecast of Trump defeat. What should have happen during the campaign for the party nomination will finally occur for the presidential election.

From above analysis I would rather state that Trump victory is still sound despite the negative polls. However to reiterate his nomination success Trump needs to modify the targeted prejudices that have been reactivated during the campaign. Before he had to address to leading frozen prejudices and beliefs which more specific to Republican oriented people and he did. Now, if keeping along the same spectrum he is doomed to fail since the shocking statement could shift $k$ from zero to values in the vicinity of $k\approx \frac{1}{2}$ but certainly not $k=1$. To reiterate the $k=1$ shift requires to activate frozen prejudices which are shared by all Americans both Democrats and Republicans. At the same time to adopt a most respectable attitude will make him lose for sure as it would have done for the Republican nomination. 

\section{Conclusion}

The sesame of the Trump campaign can be summarized as follows. By shocking the potential electors, who thus start to be against him, Trump activates frozen prejudices, which in turn decrease the threshold value at his benefit. On this basis, once the debate is launched, driven by the occurrence of collective local doubts, hostile people will shift their stand to end up supporting Trump. They do not need to endorse the shocking statement. On the contrary, they have even expressed their rejection of the statement.

In other terms Trump victory relies on both the existence of a tiny minority of openly prejudiced people and a large majority of anti-prejudice people who got somewhere hidden and locked in their values this very prejudice component, which they got in a fuzzy and unclear path during their rising childhood. 

To conclude, I have provided unanticipated Trump phenomenon with a model, the Local Majority Rule (LMR) model, which provides an explanation to his success in gaining the Republican party nomination to run for the 2016, November presidential election. Moreover the model allows a prediction about the November outcome which depends on the path Trump will reposition his shocking targets. He must activate frozen prejudices which are common to a majority of both Republican and Democrats. 

At this stage, to avoid all kinds of misleading future comments about this work, it is of importance to make a series of clarification points:

\begin{enumerate}
\item In case Donald Trump becomes the next US president, it will not be the definite proof of the rightness of the model. Indeed, it will support the legitimacy of the sociophysics approach and will demonstrate its powerful potential in the building of a hard part of social science.

\item In case Donald Trump becomes the next US president, it is of importance to state that it would not have been a yes or no prediction with fifty percent chance of success. That would dismiss the significance the model prediction. At this moment, at the end of August, 2016, according to polls and most analyses the election of Trump next November is rather improbable.

\item In case Donald Trump does not become the next US president, it will not disqualify the sociophysics approach. We are building a novel framework to explain collective phenomena in social and political sciences. A hard science is built along making predictions to check the validity of the models. Both successful and wrong predictions contribute to the elaboration of a solid and robust predictive tool. A path, which will definitively takes a long time.

\item The prediction is falsifiable and it should be to clearly enlighten the scientific nature approach of sociophysics. 

\item The numbers given and obtained to illustrate the model should not taken as precise predictions. What matters are the various trends which have been singled out.

\end{enumerate}

Last but not least, it is worth to stress that this work does not intend to express any personal political view. It aims at taking advantage of a real social event to try to discover the hidden laws, if they exist, which drive human behavior, at least part of it.

\section*{Acknowledgements}

I would like to thank Surajit Sen and Anatoly Frenkel for inviting me to give a Colloquium at the physics departments of respectively SUNY University of Buffalo (03/31/2016)  and Yeshiva University (05/4/2016). I had a chance to present my Local Majority Rule (LMR) model of opinion dynamics, which predicted the Trump nomination victory. My conclusion was received with great skepticism from both audiences.

\end{document}